\documentclass[twoside,twocolumn,9pt]{article}
\usepackage{extsizes}
\usepackage[super,sort&compress,comma]{natbib}
\usepackage[version=3]{mhchem}
\usepackage[left=1.5cm, right=1.5cm, top=1.785cm, bottom=2.0cm]{geometry}
\usepackage{balance}
\usepackage{mathptmx}
\usepackage{sectsty}
\usepackage{graphicx}
\usepackage{lastpage}
\usepackage{amsmath}
\usepackage{amssymb}
\usepackage[format=plain,justification=justified,singlelinecheck=false,font={stretch=1.125,small,sf},labelfont=bf,labelsep=space]{caption}
\usepackage{float}
\usepackage[hidelinks,draft]{hyperref}
\usepackage{fancyhdr}
\usepackage{fnpos}
\usepackage{cleveref}
\usepackage[english]{babel}
\addto{\captionsenglish}{%
  
}
\usepackage{array}
\usepackage{droidsans}
\usepackage{charter}
\usepackage[T1]{fontenc}
\usepackage[usenames,dvipsnames]{xcolor}
\usepackage{setspace}
\usepackage[compact]{titlesec}
\usepackage{xspace}

\usepackage{subcaption}
\usepackage{braket}
\usepackage{multirow}
\usepackage{bm}
\usepackage[floats=float]{chemscheme}

\crefname{figure}{Figure}{Figures}
\crefname{table}{Table}{Tables}
\crefname{equation}{Eq.}{Eqs.}
\crefname{section}{Section}{Sections}
\crefname{subsection}{Section}{Sections}

\usepackage{epstopdf}

\definecolor{cream}{RGB}{222,217,201}

\newcommand*{\kcal}{kcal mol$^{\textrm{-}1}$\xspace}

\begin{document}

\pagestyle{fancy}
\thispagestyle{plain}
\fancypagestyle{plain}{
\renewcommand{\headrulewidth}{0pt}
}

\makeFNbottom
\makeatletter
\renewcommand\LARGE{\@setfontsize\LARGE{15pt}{17}}
\renewcommand\Large{\@setfontsize\Large{12pt}{14}}
\renewcommand\large{\@setfontsize\large{10pt}{12}}
\renewcommand\footnotesize{\@setfontsize\footnotze{7pt}{10}}
\makeatother

\renewcommand{\thefootnote}{\fnsymbol{footnote}}
\renewcommand\footnoterule{\vspace*{1pt}%
\color{cream}\hrule width 3.5in height 0.4pt \color{black}\vspace*{5pt}}
\setcounter{secnumdepth}{5}

\makeatletter
\renewcommand\@biblabel[1]{#1}
\renewcommand\@makefntext[1]%
{\noindent\makebox[0pt][r]{\@thefnmark\,}#1}
\makeatother
\renewcommand{\figurename}{\small{Fig.}~}
\sectionfont{\sffamily\Large}
\subsectionfont{\normalsize}
\subsubsectionfont{\bf}
\setstretch{1.125} 
\setlength{\skip\footins}{0.8cm}
\setlength{\footnotesep}{0.25cm}
\setlength{\jot}{10pt}
\titlespacing*{\section}{0pt}{4pt}{4pt}
\titlespacing*{\subsection}{0pt}{15pt}{1pt}

\fancyfoot{}
\fancyfoot[LO,RE]{\vspace{-7.1pt}\includegraphics[height=9pt]{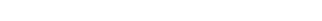}}
\fancyfoot[CO]{\vspace{-7.1pt}\hspace{11.9cm}\includegraphics{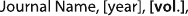}}
\fancyfoot[CE]{\vspace{-7.2pt}\hspace{-13.2cm}\includegraphics{head_foot/RF}}
\fancyfoot[RO]{\footnotesize{\sffamily{1--\pageref{LastPage} ~\textbar  \hspace{2pt}\thepage}}}
\fancyfoot[LE]{\footnotesize{\sffamily{\thepage~\textbar\hspace{4.65cm} 1--\pageref{LastPage}}}}
\fancyhead{}
\renewcommand{\headrulewidth}{0pt}
\renewcommand{\footrulewidth}{0pt}
\setlength{\arrayrulewidth}{1pt}
\setlength{\columnsep}{6.5mm}
\setlength\bibsep{1pt}

\makeatletter
\newlength{\figrulesep}
\setlength{\figrulesep}{0.5\textfloatsep}

\newcommand{\topfigrule}{\vspace*{-1pt}%
\noindent{\color{cream}\rule[-\figrulesep]{\columnwidth}{1.5pt}} }

\newcommand{\botfigrule}{\vspace*{-2pt}%
\noindent{\color{cream}\rule[\figrulesep]{\columnwidth}{1.5pt}} }

\newcommand{\dblfigrule}{\vspace*{-1pt}%
\noindent{\color{cream}\rule[-\figrulesep]{\textwidth}{1.5pt}} }

\makeatother

\twocolumn[
  \begin{@twocolumnfalse}
{\includegraphics[height=30pt]{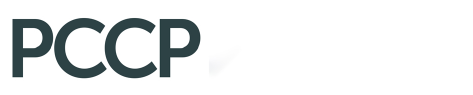}\hfill\raisebox{0pt}[0pt][0pt]{\includegraphics[height=55pt]{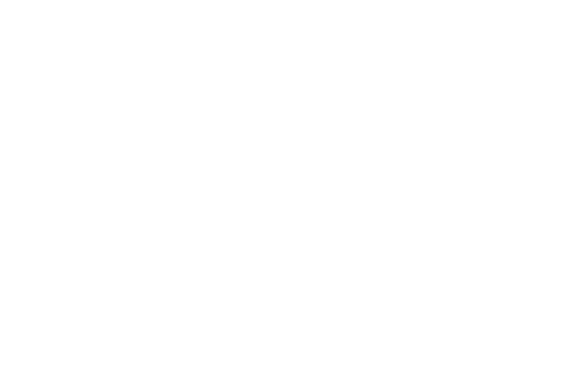}}\\[1ex]
\includegraphics[width=18.5cm]{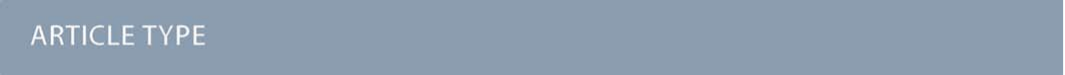}}\par
\vspace{1em}
\sffamily
\begin{tabular}{m{4.5cm} p{13.5cm} }

\includegraphics{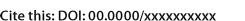} & \noindent\LARGE{\textbf{Simulating Transient X-ray Photoelectron Spectra of \ce{Fe(CO)5} and Its Photodissociation Products With Multireference Algebraic Diagrammatic Construction Theory}} \\
\vspace{0.3cm} & \vspace{0.3cm} \\

 & \noindent\large{Nicholas P.~Gaba,\textit{$^{a}$} Carlos E.~V.~de Moura,\textit{$^{a}$} Rajat Majumder,\textit{$^{a}$} and Alexander Yu.~Sokolov$^{\ast}$\textit{$^{a}$}} \\

\includegraphics{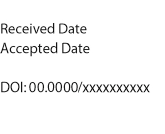} & \noindent\normalsize{
Accurate simulations of transient X-ray photoelectron spectra (XPS) provide unique opportunities to bridge the gap between theory and experiment in understanding the photoactivated dynamics in molecules and materials.
However, simulating X-ray photoelectron spectra along a photochemical reaction pathway is challenging as it requires accurate description of electronic structure incorporating core-hole screening, orbital relaxation, electron correlation, and spin--orbit coupling in excited states or at nonequilibrium ground-state geometries. 
In this work, we employ the recently developed multireference algebraic diagrammatic construction theory (MR-ADC) to investigate the core-ionized states and X-ray photoelectron spectra of \ce{Fe(CO)5} and its photodissociation products (\ce{Fe(CO)4}, \ce{Fe(CO)3})  following excitation with 266 nm light.
The simulated transient Fe 3p and CO $3\sigma$ XPS spectra incorporating spin--orbit coupling and high-order electron correlation effects are shown to be in a good agreement with the experimental measurements by Leitner et al. [{\it J.\@ Chem.\@ Phys.} {\bf 149}, 044307 (2018)].
Our calculations suggest that core-hole screening, spin--orbit coupling, and ligand-field splitting effects are similarly important in reproducing the experimentally observed chemical shifts in transient Fe 3p XPS spectra of iron carbonyl complexes. 
Our results also demonstrate that the MR-ADC methods can be very useful in interpreting the transient XPS spectra of transition metal compounds.
} \\

\end{tabular}

 \end{@twocolumnfalse} \vspace{0.6cm}

  ]

\renewcommand*\rmdefault{bch}\normalfont\upshape
\rmfamily
\section*{}
\vspace{-1cm}


\footnotetext{\textit{$^{a}$~Department of Chemistry and Biochemistry, The Ohio State University, Columbus, Ohio, 43210, USA. E-mail: vieirademoura.2@osu.edu, sokolov.8@osu.edu}}

\footnotetext{\dag~Electronic Supplementary Information (ESI) available: optimized geometries of iron carbonyl complexes, selection of active spaces, total energies for optimized geometries, summary of numerical results and spin--orbit coupling implementation. See DOI: }



\section{Introduction}

Time-resolved X-ray photoelectron spectroscopy (TR-XPS) is a versatile and highly sensitive experimental technique for probing electron and nuclear dynamics in molecules and materials.\cite{Stolow:2004p1719,Penfold:2013, Chergui:2017p11025, Ryland:2018p4691, Bhattacherjee:2018p3203}
In the TR-XPS experiment, the chemical system is sequentially irradiated with two photons: an ultraviolet (UV) or visible pump, which excites the system and initiates the dynamics, and an X-ray or extreme UV (XUV) probe that ionizes an electron out of a core orbital.
Changing the time delay between the two photons and measuring the electron count at a particular probe frequency yields the TR-XPS spectra that provide information about the electronic structure and dynamics near the localized core orbital being probed.
Advances in X-ray radiation sources and enhancements in laser technology\cite{Allaria:2012p699, Hellmann:2012p013062, Pellegrini:2016p15006, Rossbach:2019p1} have made it possible to apply TR-XPS to a wide range of problems, including investigating the charge dynamics in semiconductor interfaces,\cite{neppl_time-resolved_2015,Gessner:2016p138,borgwardt_photoinduced_2020} tracking bond dissociation processes,\cite{nugent-glandorf_ultrafast_2001,Loh:2008p204302,Leitner:2018p44307,Gabalski:2023p7126} and monitoring the excited-state dynamics of molecules.\cite{Stolow:2003p89,Liu:2020p021016,mayer_following_2022} 

As the utilization of TR-XPS continues to grow, there is an increasing demand for theoretical methodology that can help interpreting its transient spectral features.
In contrast to conventional XPS that measures the response of a chemical system in the ground electronic state near equilibrium geometries,\cite{Chen:2020p1048, Stevie:2020p63204, Kalha:2021p233001} TR-XPS probes the energies of core electrons in excited states or at nonequilibrium ground-state geometries where electronic structure may exhibit significant multiconfigurational character. 
In these situations, standard single-reference methods such as ground-state and time-dependent density functional theory,\cite{Besley:2020p1306, Hua:2020ew, Prendergast:2006p215502, Lopata:2012p3284} linear-response and equation-of-motion coupled cluster theory,\cite{Coriani:2015p181103, Liu:2019p1642, Liu:2021p1536, Halbert:2021cl, Vila:2021p} and the $GW$\cite{Golze:2018hr,Keller:2020fo} family of methods may prove to be unreliable and one may need to use multireference approaches.\cite{Rankine:2021p4276}

However, most of the available multireference methods are designed to simulate electronic excitations in frontier molecular orbitals, called the active space, and are not well-suited to describe core excitation or ionization processes.
Incorporating the core orbitals into the active space and restricting excitations in the reference wavefunction allows to overcome this problem, but introduces approximations that can be difficult to control.\cite{Agren:1993p45,Rocha.2011,Rocha.2011vjw,Josefsson:2012p3565,Moura:2013p2027,Grell:2015gu,Pinjari:2016p477,Guo:2016p3250,Montorsi:2022p1003,Roemelt:2013p204101,Corral.201794l,Bhattacharya:2021jp,Brabec:2012p171101,Huang:2022p219,Huang:2023p124112}
Alternatively, core excitations can be described using linear-response or equation-of-motion multireference theories,\cite{Yeager:1979p77,Yeager:1992p133,Nichols:1998p293,HelmichParis:2019p174121,HelmichParis:2021p26559,Kohn:2019p041106,Maganas:2019p104106} which in addition to the excitations in the active space incorporate all single excitations from non-active molecular orbitals.
Unfortunately, many of these methods are based on non-Hermitian operators or eigenvalue problems, which complicate the evaluation of excited-state properties and can produce unphysical (complex-valued) excitation energies.

Recently, we developed multireference algebraic diagrammatic construction theory (MR-ADC) for simulating excited states and electronic spectra of molecules.\cite{Sokolov:2018p204113,Chatterjee:2019p5908,Chatterjee:2020p6343,Mazin:2021p6152,Banerjee:2023p3037}
In addition to excited states in the active space, MR-ADC incorporates all single and double excitations involving non-active molecular orbitals starting with its second-order approximation (MR-ADC(2)).
The MR-ADC methods are intruder-state-free, have computational cost similar to that of conventional low-order multireference perturbation theories, are guaranteed to yield real-valued excitation energies, and can be straightforwardly adapted for the simulations of XPS spectra by utilizing the core-valence separation (CVS) technique.\cite{Cederbaum:1980p206, Barth:1981p1038,Moura:2022p8041,Moura:2022p4769,Mazin:2023p4991}
The CVS-MR-ADC methods have been used to calculate the potential energy surfaces of core-excited states in diatomic nitrogen and the ground-state XPS spectra of ozone molecule and benzyne biradicals with significant multiconfigurational character.\cite{Moura:2022p8041,Moura:2022p4769,Mazin:2023p4991}

\begin{figure}[t!]
	\centering
	\includegraphics[width=1.0\columnwidth]{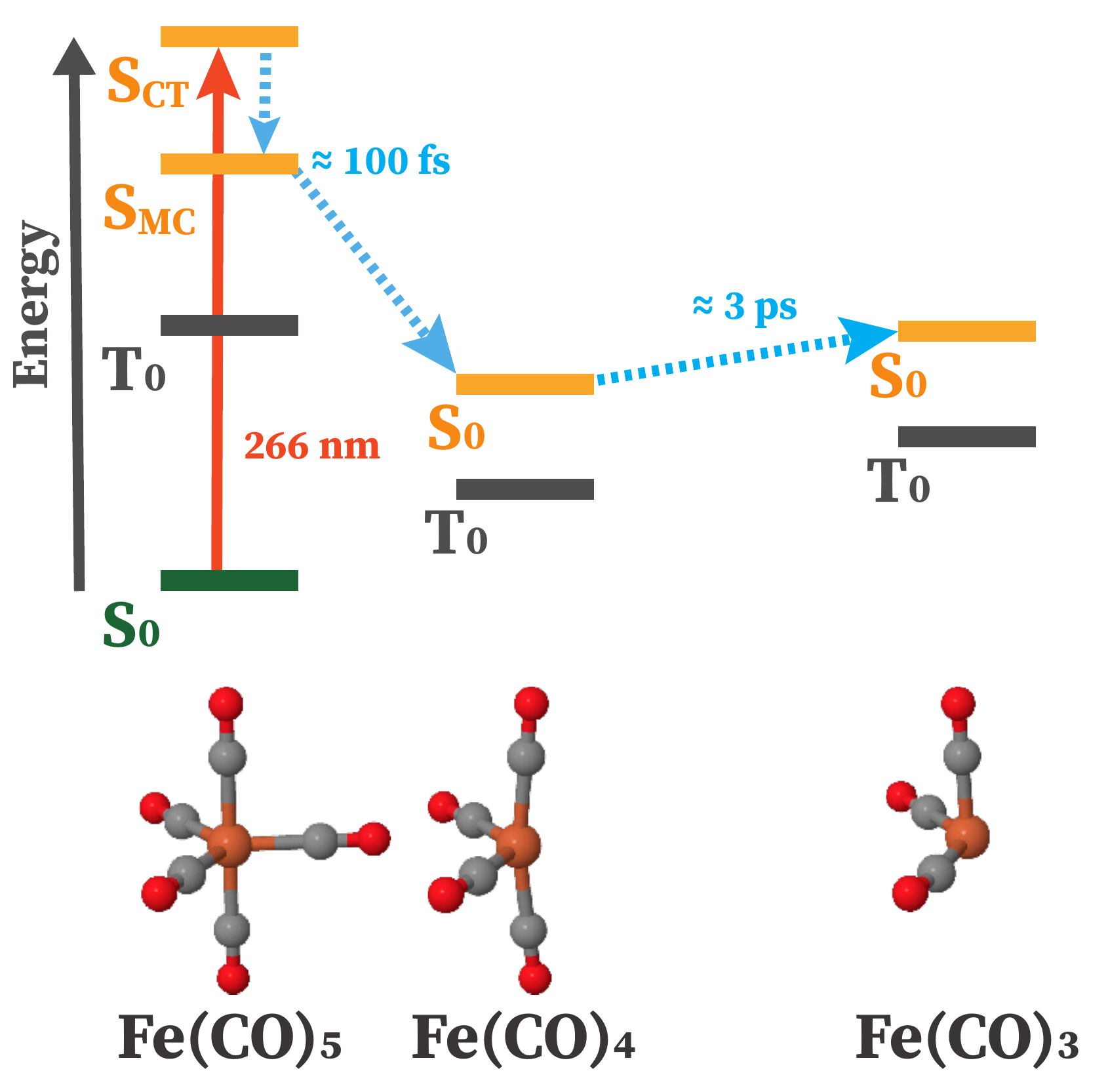}
	\caption{Schematic illustration of the singlet photodissociation pathway initiated after irradiating \ce{Fe(CO)5} with the 266 nm light in the gas phase.
	Initial excitation populating the metal-to-ligand charge-transfer state ($\mathrm{S_{CT}}$) is followed by the ultrafast internal conversion into one of the metal-centered states ($\mathrm{S_{MC}}$), which are dissociative with respect to the Fe-CO bonds.
	Subsequent internal conversion results in the formation of \ce{Fe(CO)4} and \ce{Fe(CO)3} in their lowest-energy singlet states ($\mathrm{S_{0}}$) within the next $\sim$ 100 fs and 3 ps, respectively.
	The lowest-energy triplet states ($\mathrm{T_{0}}$) are not involved in this pathway.
	}
		\label{fig:photodissociation_diagram}
\end{figure}

In this work, we present the first CVS-MR-ADC study of transient XPS spectra by applying these methods to iron pentacarbonyl (\ce{Fe(CO)5}) and its photodissociation products following the excitation with 266 nm light (\ce{Fe(CO)}$_x$, $x$ = 3, 4).
The photodissociation of \ce{Fe(CO)5} has been studied using a variety of time-resolved experimental techniques in solution and gas phase.
The solution phase dissociation has long been identified to involve triplet states of the photoproducts,\cite{portius_unraveling_2004} and a femtosecond-resolution resonant inelastic X-ray scattering study has confirmed the production of \ce{Fe(CO)4} in its triplet ground state or coordinated with solvent molecules.\cite{Wernet:2015p78}
The gas phase dissociation, however, has been the subject of debate over the specific pathway followed.
An early study using transient infrared spectroscopy showed that photoexcitation causes \ce{Fe(CO)5} to lose several of its \ce{CO} ligands, which number depends on the wavelength of excitation.\cite{Seder:1986p1977}

A particular attention has been devoted to investigating the gas-phase photodissociation dynamics of \ce{Fe(CO)5} following the excitation at 266 nm, which causes the loss of two CO ligands to form \ce{Fe(CO)3}.\cite{Trushin:2000p1997, ColeFilipiak:2021p}
Starting with \ce{Fe(CO)5} in its $^1A_{1}^\prime$ (D\textsubscript{3h} symmetry) ground state,\cite{daniel_theoretical_1984, Ihee:2001p1532, Pierloot:2003p2083} absorption at 266 nm can populate several closely lying states,\cite{daniel_theoretical_1984, Persson:1994p6810} of which singlet excited $E^\prime$ and $A_{2}^{\prime \prime}$  have been identified as potentially participating in the dissociation pathway.\cite{Rubner:1999p489, Trushin:2000p1997, Ihee:2001p1532, Malcomson:2019p825, ColeFilipiak:2021p}
Trushin et al.\cite{Trushin:2000p1997}~proposed a completely singlet dissociation pathway proceeding through the C\textsubscript{2v}-symmetric $^1A\textsubscript{1}$ state of \ce{Fe(CO)4} on a timescale of less than 100\,fs\cite{Poliakoff:2001p2809}, followed by subsequent dissociation into singlet \ce{Fe(CO)3} within the next 3.3\,ps.
This singlet pathway, illustrated in \cref{fig:photodissociation_diagram}, has been supported by studies utilizing ultrafast electron diffraction,\cite{Ihee:2001p1532} TR-XPS,\cite{Wernet:2017p211103,Leitner:2018p44307} and time-resolved XUV absorption in conjunction with quantum chemical calculations.\cite{tross_femtosecond_2023}
In particular, the gas-phase TR-XPS spectra measured by Wernet et al.\@ and Leitner et al.\@ provide us with the opportunity to benchmark the accuracy of CVS-MR-ADC in simulating transient spectra of transition metal complexes such as \ce{Fe(CO)5}. 

This paper is organized as follows.
We first briefly describe the theory behind CVS-MR-ADC (\cref{sec:Theory}) and outline computational details (\cref{sec:ComputationalDetails}).
In \cref{sec:Results:XPS_3p,sec:Results:XPS_3sigma}, we present the simulated transient XPS spectra for the core-level Fe 3p and inner-valence CO $3 \sigma$  ionizations, respectively, and compare them to the experimental measurements in the XUV region of electromagnetic spectrum from Leitner et al.\cite{Leitner:2018p44307}
We analyze the results of our calculations in \cref{sec:Results:comp_parameters} where we investigate the effect of molecular geometries, spin--orbit coupling effects, active space, and basis set on the simulated XPS spectra.
We present our conclusions  in \cref{sec:Conclusions}.

\section{Theory}
\label{sec:Theory}

\subsection{Multireference algebraic diagrammatic construction theory}
\label{sec:Theory:mr_adc}

We start with a brief overview of MR-ADC for ionization energies. 
A more detailed presentation of the theory can be found elsewhere.\cite{Banerjee:2023p3037, Sokolov:2018p204113, Chatterjee:2019p5908, Chatterjee:2020p6343, Moura:2022p8041, Moura:2022p4769}

The principal information about ionization of valence and core orbitals is contained in the one-particle Green's function ($G_{pq}(\omega)$)\cite{dickhoff2008many,fetter2012quantum} that describes the linear response of a chemical system in the electronic state $\ket{\Psi}$ with energy $E$ to the ionizing radiation with frequency $\omega$:
\begin{equation}
	\label{eq:greens_function}
	G_{pq}(\omega) = \langle \Psi | a_{p}^{\dag} (\omega - H - E)^{-1} a_{q} | \Psi \rangle,
\end{equation}
Here, $a_{q}$ is the annihilation operator that removes an electron from molecular orbital $\psi_q$, $H$ is the Hamiltonian of the system, and $a_{p}^\dag$ is the creation operator that probes the occupancy of $\psi_p$ by adding an electron.
The Green's function can be used to compute the density of states as a function of $\omega$
\begin{equation}
	\label{eq:dos}
	A(\omega) = - \frac{1}{\pi} \mathrm{Im}[\mathrm{Tr} \; \mathbf{G}(\omega)]
\end{equation}
which to a good approximation represents the photoelectron spectrum of the system.

MR-ADC simulates the photoelectron spectra by computing efficient multireference approximations to $G_{pq}(\omega)$ that are constructed by i) selecting a subset of frontier molecular orbitals, which give rise to the multiconfigurational character of $\ket{\Psi}$, as an active space, ii) computing a complete active space self-consistent field (CASSCF) wavefunction\cite{Werner.1980,Werner.1981,Knowles.1985} for the selected active space ($\ket{\Psi_0}$), and iii) expanding $G_{pq}(\omega)$ in the multireference perturbation series starting with $\ket{\Psi_0}$ as the zeroth-order wavefunction and dividing the Hamiltonian $H$ into the zeroth-order ($H^{(0)}$) and perturbation ($V = H - H^{(0)}$) contributions. 
Choosing $H^{(0)}$ as the Dyall Hamiltonian,\cite{Dyall:1995p4909,sokolov2024multireference} the $n$th-order approximation to $G_{pq}(\omega)$ (termed as MR-ADC($n$)) is constructed by representing this function in the nondiagonal matrix form
\begin{equation}
	\label{eq:gf_adc}
	\mathbf{G}(\omega) = \mathbf{T} (\omega \mathbf{S} - \mathbf{M})^{-1} \mathbf{T}^{\dag}
\end{equation}
and expanding each matrix in this expression up to the order $n$.
Due to the two-electron form of the Dyall Hamiltonian, the MR-ADC approximations do not suffer from intruder-state problems and do not require using level shift parameters, as opposed to complete and restricted active-space perturbation theories\cite{Andersson:1990p5483,Andersson:1992p1218,Roos:1995p215,Finley:1998p299} that employ a one-electron $H^{(0)}$.

The energies of ionized states ($\boldsymbol{\Omega}$) are computed by solving the generalized eigenvalue problem for the effective Hamiltonian matrix $\mathbf{M}$
\begin{equation}
	\label{eq:mradc-eigenvalue-problem}
	\mathbf{M} \mathbf{Y} = \mathbf{S} \mathbf{Y} \boldsymbol{\Omega}
\end{equation}
in the basis of nonorthogonal excitations with overlap $\mathbf{S}$.
Combining the eigenvectors $\mathbf{Y}$ with the effective transition moments matrix $\mathbf{T}$ allows to compute the spectroscopic amplitudes 
\begin{equation}
	\label{eq:spec_amplitudes}
	\mathbf{X} = \mathbf{T} \mathbf{S}^{-1/2} \mathbf{Y}
\end{equation}
which contain information about the intensities of transitions in photoelectron spectra.

Two MR-ADC approximations are employed in this work: the strict second-order (MR-ADC(2)) and extended second-order (MR-ADC(2)-X) methods.
Both MR-ADC(2) and MR-ADC(2)-X incorporate all electronic excitations in active orbitals and up to double excitations involving at least one non-active orbital. 
The MR-ADC(2) method approximates each matrix in \cref{eq:gf_adc} up to the second order in multireference perturbation theory.
The MR-ADC(2)-X approach improves the description of orbital relaxation effects by incorporating the third-order contributions to $\mathbf{M}$ and $\mathbf{T}$ in the treatment of double excitations.\cite{Banerjee:2023p3037, Chatterjee:2020p6343, Moura:2022p8041, Moura:2022p4769, sokolov2024multireference}
The ability to describe excitations in non-active orbitals contrasts MR-ADC with conventional multireference perturbation theories that are restricted to exciting electrons only in the active space.

\subsection{Core-valence separation}
\label{sec:Theory:cvs}

The core-ionized states probed in X-ray photoelectron spectroscopy (XPS) correspond to the high-energy excited states that are deeply embedded in the eigenspectrum of MR-ADC effective Hamiltonian matrix $\mathbf{M}$ (\cref{eq:mradc-eigenvalue-problem}) and are difficult to extract using conventional eigenvalue solvers. 
Fortunately, due to their high localization in space and energy spectrum, the coupling of core-ionized wavefunctions with other electronic states is usually rather weak and can be neglected, which is known as the core-valence separation (CVS) approximation.\cite{Cederbaum:1980p206, Barth:1981p1038}
The CVS approximation has been shown to produce small (often negligible) systematic errors in excitation energies and transition intensities\cite{Peng:2019p1840,Banerjee:2019p224112,Liu:2019p1642,Herbst:2020p054114,HelmichParis:2021p26559} while preventing issues with the discretization of continuum states when using incomplete atom-centered basis sets.\cite{Liu:2019p1642,Nanda:2020p141104}

Here, we employ our CVS implementation of  MR-ADC(2) and MR-ADC(2)-X to efficiently simulate core-ionized states and XPS spectra of \ce{Fe(CO)5} and its photodissociation products.\cite{Moura:2022p8041, Moura:2022p4769}
Introducing CVS in MR-ADC(2) and MR-ADC(2)-X leads to a significant reduction in computational cost as only excitations involving at least one core orbital are included in the representation of MR-ADC matrices.
In addition to core-ionized states, the utility of CVS technique has been demonstrated in MR-ADC calculations of core-excited states and X-ray absorption spectra.\cite{Mazin:2021p6152}

\subsection{Relativistic effects}
\label{sec:Theory:so_coupling}

In addition to electron correlation, core-hole screening, and orbital relaxation, simulating core-excited and core-ionized states requires accurate treatment of relativistic effects.\cite{Lee:2010p9715,Kasper:2018p1998,Maganas:2019p104106,Stetina:2019p234103,Vidal:2020p8314} 
Among the two major relativistic interactions are i) scalar relativistic effects due to the contraction or expansion of atomic orbitals that shift the energies of transitions in XPS spectra and ii) spin--orbit coupling that alters the densities of states and photoelectron spectra when exciting electrons from core orbitals with non-zero angular momentum (e.g., 2p, 3p, 3d).\cite{Pyykko:2012p45}

In this work, we use the spin-free X2C Hamiltonian\cite{Dyall.2001.4,Liu.2009} to accurately and efficiently treat scalar relativistic effects in the CASSCF and MR-ADC calculations.
To incorporate spin--orbit coupling, we employ a composite approach where the spin--orbit corrections to the MR-ADC(2) and MR-ADC(2)-X core ionization energies are computed at the MR-ADC(1) level of theory using the Breit--Pauli spin--orbit mean-field Hamiltonian.
The details of our spin--orbit MR-ADC(1) implementation are provided in the ESI.$\dag$

\section{Computational Details}
\label{sec:ComputationalDetails}

\begin{figure*}[t!]
	\centering
	\includegraphics[width=2.0\columnwidth]{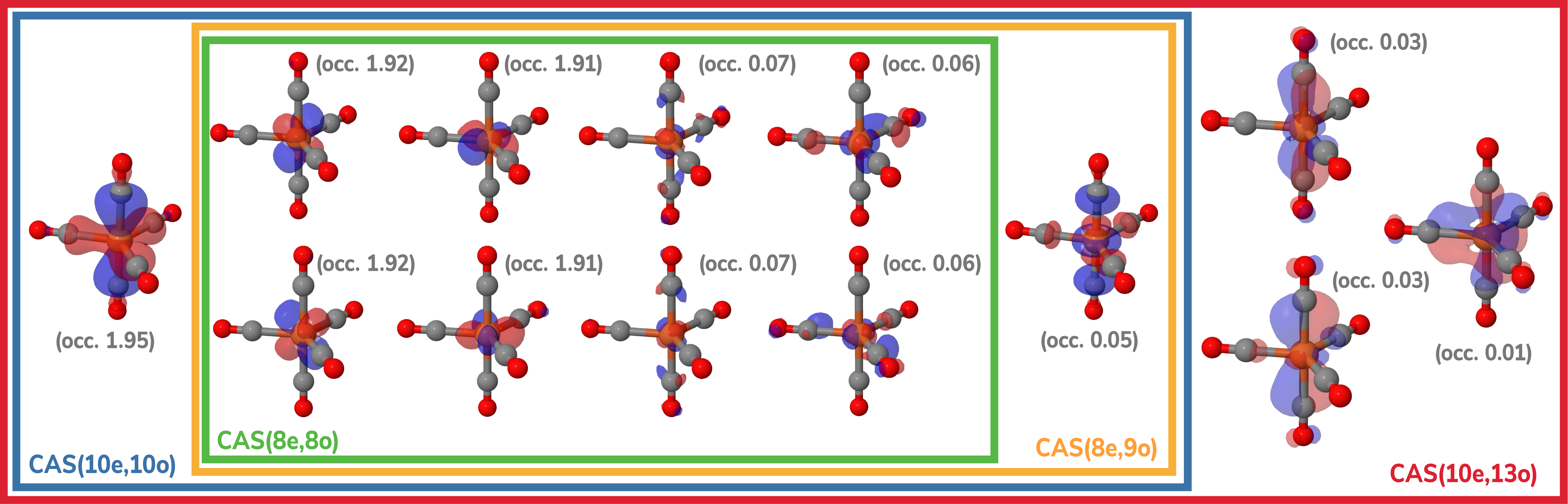}
	\caption{Active spaces of the \ce{Fe(CO)5} complex used for the multireference calculations in this work. For each orbital, natural occupations from the CAS(10e, 13o) calculation are shown in parentheses.}
		\label{fig:FeCO5_CAS}
\end{figure*}

The molecular geometries of Fe(CO)$_x$ ($x$ = 3 -- 5) were optimized using second-order N-electron valence perturbation theory (NEVPT2)\cite{Angeli:2001p10252,Angeli:2002p9138} starting with complete active space self-consistent field (CASSCF) reference wavefunctions that included 10 electrons in 10 active orbitals (CAS(10e, 10o)).
These geometry optimizations were performed for the lowest-energy singlet state of each complex.
For Fe(CO)$_5$, the active space included five Fe 3d and five CO orbitals as shown in \cref{fig:FeCO5_CAS}, in agreement with the previous multireference studies of this molecule.\cite{Persson:1994p6810, Pierloot:2003p2083, Tsuchiya:2006p1123}
For \ce{Fe(CO)4} and \ce{Fe(CO)3}, CAS(10e, 10o) incorporated five Fe 3d and five CO-based orbitals.
To simulate experimental conditions with free CO present,\cite{Leitner:2018p44307} uncoordinated CO molecules were incorporated into the \ce{Fe(CO)4} and \ce{Fe(CO)3} calculations.
All geometry optimizations were performed with the Molpro program\cite{Werner:2012p242, Werner:2020p144107, MOLPRO_2021} using the def2-QZVPP basis set.\cite{Weigend:2005p3297} 
Structures with free CO were optimized by constraining the \ce{Fe-C} distance at 10.0 \AA and allowing all other structural parameters to relax.
Additionally, two-electron integrals were approximated using density fitting\cite{Whitten:1973p4496,Dunlap:1979p3396,Vahtras:1993p514,Feyereisen:1993p359} with the def2-QZVPP-JKFIT and def2-QZVPP-RI auxiliary basis sets employed for CASSCF and NEVPT2, respectively.\cite{Weigend:1998p143,Hattig:2004p59,Hellweg:2007p587,Weigend:2008p167} 

The XPS spectra of singlet Fe(CO)$_x$ ($x$ = 3 -- 5) {for} the {core-level} Fe 3p  and {inner-valence} \ce{CO} 3$\sigma$  ionization were computed using the CVS-MR-ADC(2) and CVS-MR-ADC(2)-X implementations in the \textsc{Prism} program.\cite{prism}
The \textsc{Prism} code was interfaced with the \textsc{Pyscf} package,\cite{Sun:2020p024109} which provided the one- and two-electron integrals and CASSCF reference wavefunctions.
For brevity, we refer to CVS-MR-ADC(2) and CVS-MR-ADC(2)-X as MR-ADC(2) and MR-ADC(2)-X henceforth. 

The MR-ADC calculations were performed at the NEVPT2 singlet optimized geometries using the CASSCF reference wavefunctions with the CAS(10e, 13o) active space (\cref{fig:FeCO5_CAS}) and fully uncontracted def2-TZVPP basis set (unc-def2-TZVPP).
The $\eta_s = 10^{-5}$  and $\eta_d = 10^{-10}$ parameters were specified to remove linearly dependent semiinternal and double excitations, respectively.
We also considered alternative selections of equilibrium geometries, active spaces, and basis sets in \cref{sec:Results:geometry,sec:Results:active_space,sec:Results:basis_set}.
For the planar (D\textsubscript{3h}) geometry of \ce{Fe(CO)3}, a smaller CAS(8e, 12o) was used, due to a weaker interaction between the ligand $\sigma$ orbitals and Fe d$_{z^{2}}$. 

The MR-ADC simulations of XPS spectra were performed using density fitting where for each main basis set we employed the corresponding JKFIT and RI auxiliary basis sets in the CASSCF and MR-ADC calculations, respectively.\cite{Weigend:1998p143,Hattig:2004p59,Hellweg:2007p587,Weigend:2008p167}
Additionally, relativistic effects were incorporated using the spin-free X2C Hamiltonian\cite{Dyall.2001.4,Liu.2009} and spin--orbit coupling calculations as described in \cref{sec:Theory:so_coupling} and ESI.$\dag$
The XPS spectra were simulated by plotting the density of states 
\begin{align}
	\label{eq: PES}
	A(\omega) &= -\frac{1}{\pi} \mathrm{Im} \left[ \sum_{\mu}\frac{P_{\mu}}{\omega - \omega_{\mu} + i\eta} \right]
\end{align}
where $\omega_{\mu}$ denotes the MR-ADC core ionization energies, $\eta$ is an artificial broadening parameter set to 1.8 eV (similar to experimental bandwidth of 1.5 eV),\cite{Leitner:2018p44307} and $P_{\mu}$ are the MR-ADC spectroscopic factors
\begin{align}
	\label{eq:spec_factors}
	P_{\mu} = \sum_{p} |X_{p\mu}|^2
\end{align}
computed using the spectroscopic amplitudes defined in \cref{eq:spec_amplitudes}.
The difference XPS spectra of the lowest-energy singlet \ce{Fe(CO)4} and \ce{Fe(CO)3} were computed by subtracting their densities of states with that of ground-state (singlet) \ce{Fe(CO)5}.
The experimental difference spectra measured with respect to the ground-state \ce{Fe(CO)5} were obtained by digitizing the data presented in Ref.\@ \citenum{Leitner:2018p44307} using WebPlotDigitizer.\cite{Rohatgi2022}

\section{Results and Discussion}
\label{sec:Results}

\subsection{X-ray photoelectron spectra at the Fe 3p edge}
\label{sec:Results:XPS_3p}

\begin{figure}[t!]
	\centering
	\includegraphics[width=1.0\columnwidth]{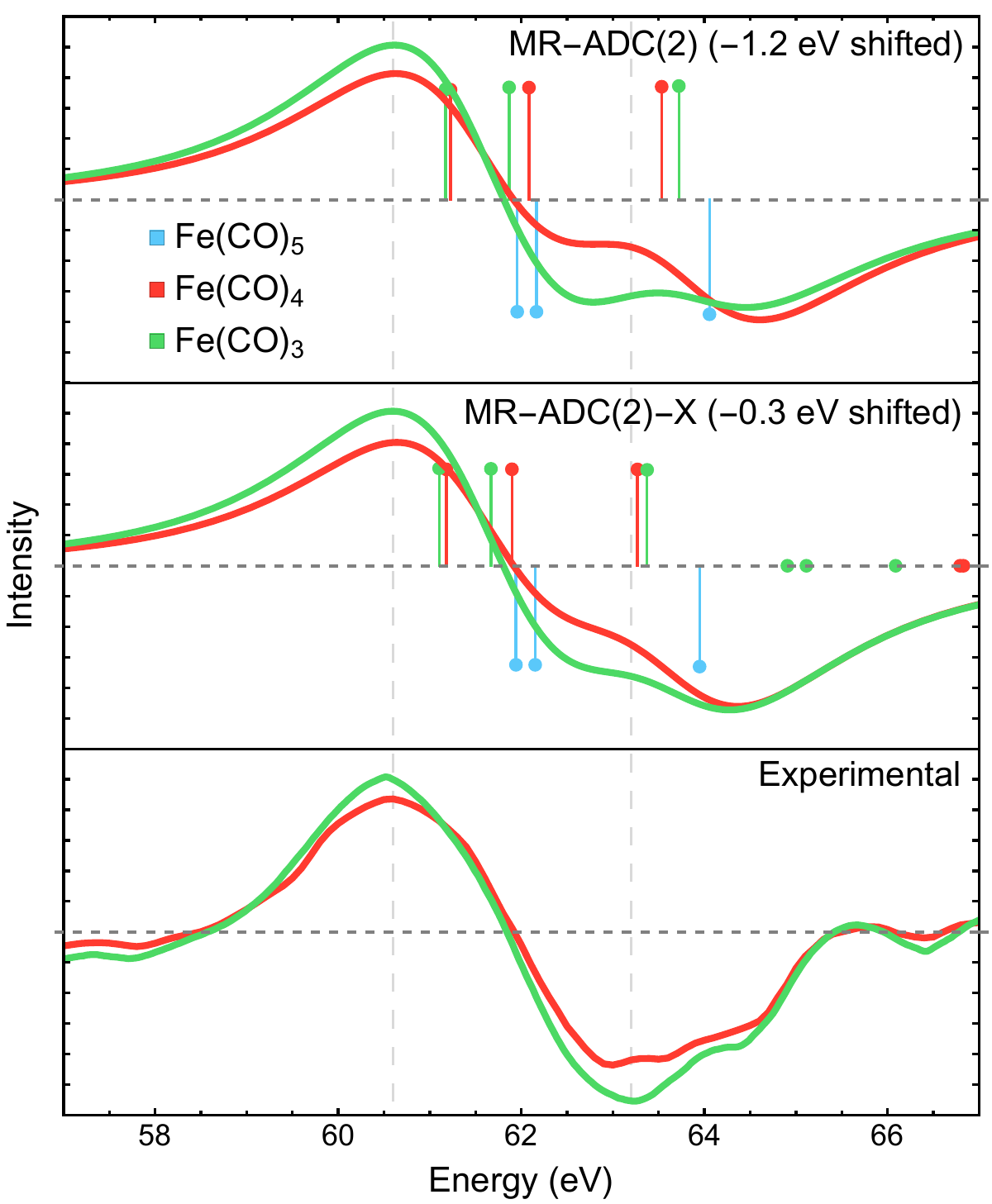}
	\caption{
	Difference \ce{Fe} 3p X-ray photoelectron spectra of \ce{Fe(CO)4} and \ce{Fe(CO)3} relative to \ce{Fe(CO)5} measured experimentally\cite{Leitner:2018p44307}  and simulated using MR-ADC(2) and MR-ADC(2)-X.
	Vertical lines represent the computed core ionization energies and the corresponding photoelectron intensities.
	The simulated spectra were shifted to align the position of zero-intensity intercept with the experimental spectrum.
	Calculations employed the CAS(10e, 13o) active space and uncontracted def2-TZVPP basis set.
	See \cref{sec:ComputationalDetails} for more computational details.
	}
	\label{fig:fe3p_diffspec}
\end{figure}

\begin{figure}[t!]
	\centering
	\includegraphics[width=1.0\columnwidth]{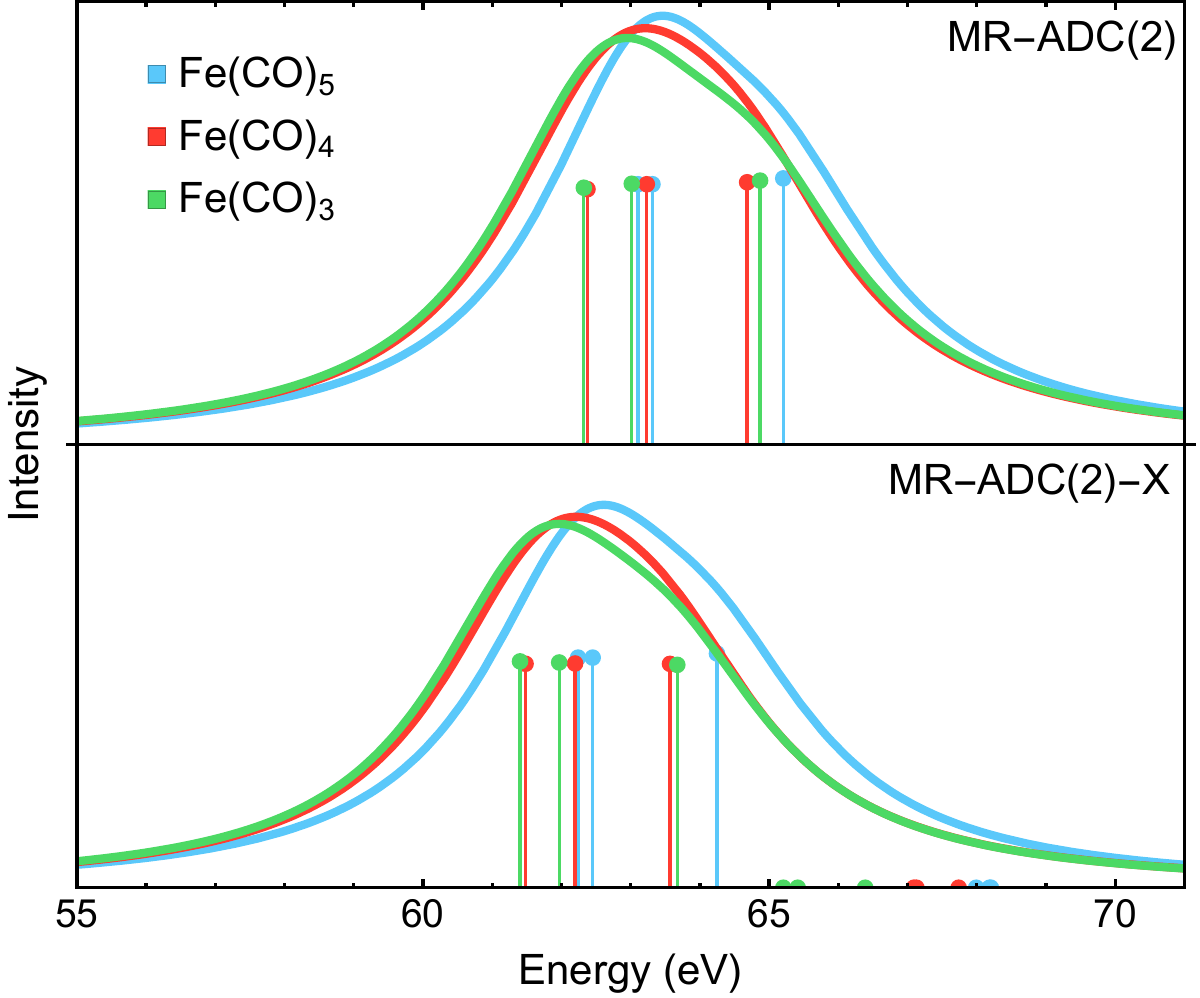}
	\caption{
	\ce{Fe} 3p X-ray photoelectron spectra of  \ce{Fe(CO)5}, \ce{Fe(CO)4}, and \ce{Fe(CO)3} simulated using MR-ADC(2) and MR-ADC(2)-X.
	Vertical lines represent the computed core ionization energies and the corresponding photoelectron intensities.
	Calculations employed the CAS(10e, 13o) active space and uncontracted def2-TZVPP basis set.
	See \cref{sec:ComputationalDetails} for more computational details.
	}
	\label{fig:fe3p_spec}
\end{figure}

\cref{fig:fe3p_diffspec} compares the transient XPS spectra of \ce{Fe(CO)4} and \ce{Fe(CO)3} simulated using MR-ADC(2) and MR-ADC(2)-X with the experimental difference spectra\cite{Leitner:2018p44307} for the core-level Fe 3p ionization.
The experimental spectrum of \ce{Fe(CO)4} shows two broad features, a negative peak at 63.2 eV and a positive peak at 60.6 eV, indicating a significant redshift in the energy of Fe 3p core-ionized states (Fe 3p$^{-1}$) upon the loss of the first CO ligand.
The dissociation of second CO ligand resulting in \ce{Fe(CO)3} does not noticeably change the positions of both features but increases their intensities by $\sim$ 15 to 20 \%.

The MR-ADC(2) and MR-ADC(2)-X simulated Fe 3p difference spectra show a good agreement with the experimental results, reproducing the magnitude of observed spectral changes.
Using the unc-def2-TZVPP basis set, the CAS(10e, 13o) active space, and a broadening parameter of 1.8 eV, the MR-ADC(2) and MR-ADC(2)-X spectra are systematically shifted by $\sim$ 1.2 and 0.3 eV, respectively, relative to experiment.
Aside from the systematic shift and the shape of negative feature, the MR-ADC(2) and MR-ADC(2)-X spectra are similar to each other and the experimental measurements.
Both methods reveal structure in the negative feature observed in the experiment, but do not correctly describe its shape. 
As shown in \cref{fig:fe3p_diffspec}, the structure of the negative feature in the simulated spectra can be attributed to a significant overlap of Fe 3p$^{-1}$ density of states between \ce{Fe(CO)5} and its photodissociation products, which can be sensitive to small changes in molecular geometry and approximations in simulations.

Careful analysis of \cref{fig:fe3p_diffspec} reveals that the changes in the energies of individual Fe 3p$^{-1}$ eigenstates upon photodissociation are significantly smaller ($\lesssim$ 1 eV) than the separation between the maxima of positive and negative features observed in the experimental and simulated difference XPS spectra ($>$ 2 eV). 
Indeed, the simulated Fe 3p XPS spectra of each iron carbonyl complex shown in \cref{fig:fe3p_spec} exhibit at most 1 eV redshift, suggesting that the energetic stabilization of Fe 3p$^{-1}$ states due to the enhanced core-hole screening upon the loss of CO ligands is significantly smaller than what can be inferred from the analysis of transient spectra.
Similar redshift ($\lesssim$ 1 eV) in the energies of Fe 3p$^{-1}$ states has been observed in the results of multiconfigurational self-consistent field (MCSCF) calculations carried out by Leitner et al.\cite{Leitner:2018p44307} who suggested that neglecting excited-state structural relaxation may be responsible for the smaller-than-expected redshift in the simulations.

Although our calculations did not incorporate vibrational dynamics and excited-state structural relaxation effects, \cref{fig:fe3p_diffspec} indicates that, in addition to $\lesssim$ 1 eV stabilization of Fe 3p$^{-1}$ states upon photodissociation, at least two more electronic factors need to be considered to explain the $>$ 2 eV separation between the features in the difference spectra: (i) strong spin--orbit coupling that splits the Fe 3p$^{-1}$ energy levels by $\sim$ 1.0 to 1.2 eV (see \cref{sec:Results:SO_coupling}) and (ii) increase in ligand-field splitting of Fe 3p orbitals in \ce{Fe(CO)4} and \ce{Fe(CO)3} compared to higher symmetry \ce{Fe(CO)5} that separates the 3p$^{-1}$ states further apart.
The small $\lesssim$ 1 eV redshift in Fe 3p$^{-1}$ states is also consistent with zero net change in the formal oxidation state of the Fe atom during the photodissociation, which is further supported by the results of Mulliken charge analysis that shows a marginal decrease in the Fe atom charge from +0.52 in \ce{Fe(CO)5} to +0.49 in \ce{Fe(CO)4} and \ce{Fe(CO)3} at the CASSCF(10e, 10o) level of theory. 

In addition to the energies of Fe 3p$^{-1}$ eigenstates, the MR-ADC calculations provide access to transition intensities, which incorporate electron correlation effects.
\cref{fig:fe3p_diffspec} demonstrates that MR-ADC(2) and MR-ADC(2)-X accurately capture the $\sim$ 15 to 20 \% increase in spectral intensities observed in experiment during the loss of second CO ligand.
In our simulations, this intensity enhancement can be traced back to the difference in spread of Fe 3p$^{-1}$ states, which is larger in \ce{Fe(CO)3} compared to \ce{Fe(CO)4} by $\sim$ 0.2 eV.
In addition, the photoelectron transitions in \ce{Fe(CO)3} exhibit a small increase in computed spectroscopic intensities ($\sim$ 1\%), which provides a minor contribution to the observed intensity enhancement.

\subsection{X-ray photoelectron spectra at the CO 3$\sigma$ edge}
\label{sec:Results:XPS_3sigma}

\begin{figure}[t!]
	\centering
	\includegraphics[width=1.0\columnwidth]{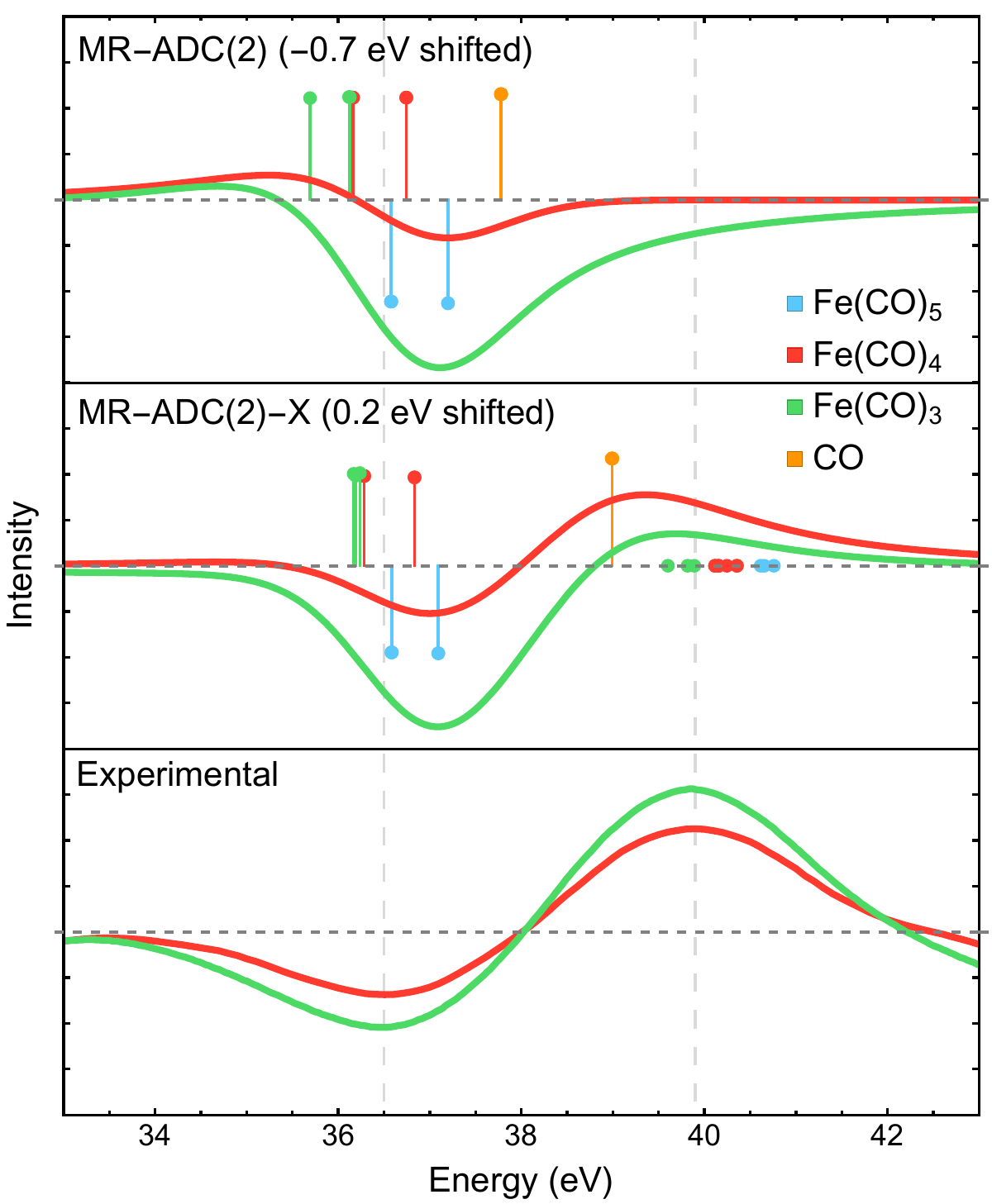}
	\caption{
		Difference CO 3$\sigma$$^{-1}$ X-ray photoelectron spectra of \ce{Fe(CO)4} and \ce{Fe(CO)3} relative to \ce{Fe(CO)5} measured experimentally\cite{Leitner:2018p44307}  and simulated using MR-ADC(2) and MR-ADC(2)-X.
		Vertical lines represent the computed core ionization energies and the corresponding photoelectron intensities.
		The positive feature in experimental spectrum corresponds to free CO.
		To simulate free CO being present, one uncoordinated CO ligand was included in the  \ce{Fe(CO)4} and \ce{Fe(CO)3} computations. 
		The simulated spectra were shifted as indicated on each plot.
		Calculations employed the CAS(10e, 13o) active space and uncontracted def2-TZVPP basis set.
		See \cref{sec:ComputationalDetails} for more computational details.
	}
	\label{fig:co3sigma_diffspec}
\end{figure}

The XPS difference spectra {for} the \ce{CO} 3$\sigma$ {inner-valence ionization} measured in experiment\cite{Leitner:2018p44307} and simulated with MR-ADC are shown in \cref{fig:co3sigma_diffspec}.
The experimental spectra exhibit a negative peak at $\sim$ 36.5 eV and a positive feature at $\sim$ 40 eV, which were assigned to the loss of coordinated CO in \ce{Fe(CO)5} and the generation of free CO in the reaction environment, respectively. 
In addition, the analysis of experimental data suggested that the \ce{CO} ligands bound in \ce{Fe(CO)4} and \ce{Fe(CO)3} show a redshift in 3$\sigma$ binding energy, but the precise value of the redshift could not be established due to poor resolution.\cite{Leitner:2018p44307}

The MR-ADC calculations for \ce{Fe(CO)4} and \ce{Fe(CO)3} were performed with one uncoordinated CO molecule to simulate the presence of free CO.
In agreement with experiment, the free CO molecule has a higher 3$\sigma$ binding energy relative to CO in \ce{Fe(CO)5}, with a $\sim$ 2.3 eV blueshift calculated at the MR-ADC(2)-X level of theory.
Dissociating CO ligands results in a small ($\sim$ 0.5 eV) redshift in the 3$\sigma$ binding energy of coordinated CO, which is not sufficiently large to produce a positive feature with significant intensity when using the broadening parameter of 1.8 eV.
These results are consistent with experimental data where a positive feature due to coordinated CO was not observed within the instrument resolution.\cite{Leitner:2018p44307} 

\subsection{Analysis of Computational Results}
\label{sec:Results:comp_parameters}

In the following, we will investigate how various parameters affect the result of MR-ADC simulations performed in this work, including the effect of equilibrium geometries, spin--orbit coupling, selection of active space and basis set, and the role of multireference treatment.
In our analysis, we will focus on the Fe 3p edge, which was studied with a higher resolution in the experiment.\cite{Leitner:2018p44307}
The results for CO 3$\sigma$ edge can be found in the ESI.$\dag$

\subsubsection{Equilibrium Geometries}
\label{sec:Results:geometry}

\begin{table}[t!]
	\centering
	\caption{Equilibrium dissociation energies ($D_e$, \kcal) of iron carbonyl complexes in their lowest singlet electronic states computed relative to \ce{Fe(CO)5} using the NEVPT2 method.
	The calculations were performed using the def2-TZVPP basis set and CAS(10e, 10o) active space.}
	\begin{tabular}{lc}
		\hline \hline
		Complex & $D_e$, \kcal \\
		\hline
		\ce{Fe(CO)5}& 0.0 \\
		\ce{Fe(CO)4} (C\textsubscript{2v}) + CO   & 52.2\\
		\ce{Fe(CO)4} (C\textsubscript{3v}) + CO  & 58.6\\
		\ce{Fe(CO)3} (C\textsubscript{s}) + 2CO  & 99.8\\
		\ce{Fe(CO)3} (D\textsubscript{3h}) + 2CO  & 137.3\\
		\hline \hline
	\end{tabular}
	\label{tbl:energies}
\end{table}

\begin{figure}[t!]
	\centering
	\includegraphics[width=1.0\columnwidth]{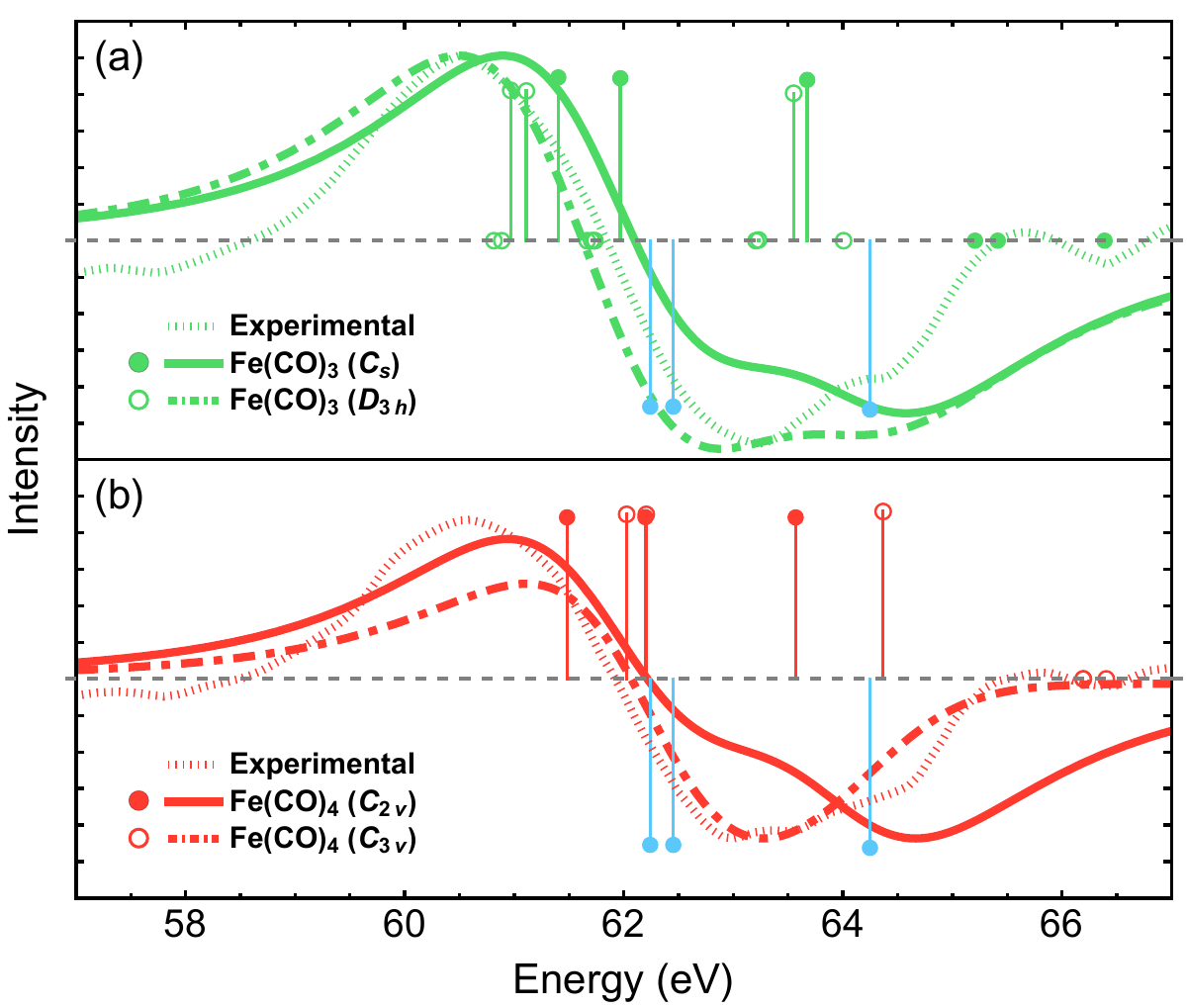}
	\caption{
	Difference \ce{Fe} 3p X-ray photoelectron spectra for the two lowest-energy structures of \ce{Fe(CO)3} (a) and \ce{Fe(CO)4} (b) simulated using MR-ADC(2)-X and compared to the experimental spectrum from Ref.\@ \citenum{Leitner:2018p44307}.
	Vertical lines represent the computed core ionization energies and the corresponding photoelectron intensities.
	Blue lines correspond to \ce{Fe(CO)5}.
	No shift was applied to the simulated spectra.
	See \cref{sec:ComputationalDetails} for computational details.
	}		
	\label{fig:fe3p_isomers}
\end{figure}

The \ce{Fe(CO)4} and \ce{Fe(CO)3} complexes are known to have two bound minima on their potential energy surfaces that can participate in the photodissociation dynamics.\cite{Trushin:2000p1997,tross_femtosecond_2023}
Dissociation energies ($D_e$) of these isomers relative to \ce{Fe(CO)5} computed using the NEVPT2 method with the CAS(10e, 10o) active space are reported in \cref{tbl:energies}.

Two conformations of singlet \ce{Fe(CO)4} can be produced following the dissociation of either the equatorial or axial CO ligands in \ce{Fe(CO)5}, resulting in the C\textsubscript{2v} or C\textsubscript{3v} structures, respectively. 
The two isomers have similar energies, with the C\textsubscript{2v} structure being favored by only 6.4 \kcal at the NEVPT2 level of theory (\cref{tbl:energies}).
Recent density functional theory study by Tross et al.\cite{tross_femtosecond_2023} suggested that \ce{Fe(CO)4} may spontaneously interconvert between the C\textsubscript{2v} and C\textsubscript{3v} structures during the course of photodissociation dynamics. 
The simulated difference spectra for the C\textsubscript{2v} and C\textsubscript{3v} isomers are shown in \cref{fig:fe3p_isomers}(b) and are compared to the experimental results. 
When convoluted with the 1.8 eV broadening, the C\textsubscript{3v} spectrum exhibits a smaller redshift of the Fe 3p feature compared to that for the C\textsubscript{2v} isomer.
Adding the two difference spectra together with 1:1 ratio provides a better agreement with the experimental results than the spectrum of each isomer alone, which may support the findings of Tross et al.\@ that both structures are similarly important during the course of photodissociation, and warrants further investigations that incorporate nuclear dynamics effects.

The dissociation of second CO ligand can result in two \ce{Fe(CO)3} structures with the C\textsubscript{s} and D\textsubscript{3h} symmetries.
As shown in \cref{tbl:energies}, the C\textsubscript{s} structure is lower in energy by 37.5 \kcal and is therefore expected to be energetically preferred. 
\cref{fig:fe3p_isomers}(a) shows that both structures exhibit similar redshifts of Fe 3p band.
The spectrum of D\textsubscript{3h} isomer has a slightly broader negative feature, which is less consistent with the experiment compared to that of C\textsubscript{s} structure, given our simulation parameters.
However, overall, these results suggest that distinguishing spectral features of the C\textsubscript{s} and D\textsubscript{3h} isomers would require achieving a higher resolution of experimental measurements if both structures participated in the photodissociation dynamics.

\subsubsection{Role of Spin-Orbit Coupling Effects}
\label{sec:Results:SO_coupling}

\begin{figure}[t!]
	\centering
	\includegraphics[width=1.0\columnwidth]{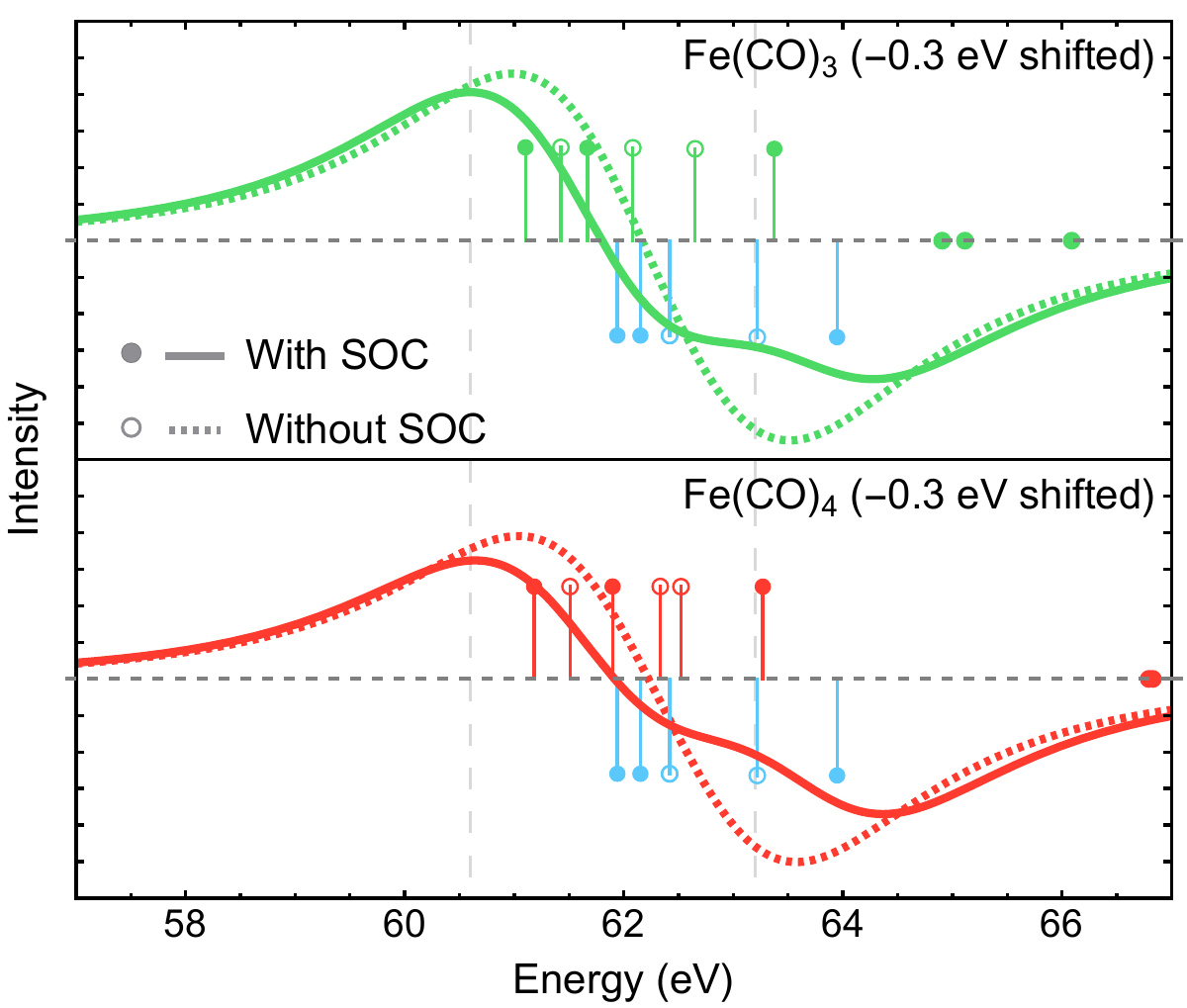}
	\caption{
	Difference \ce{Fe} 3p X-ray photoelectron spectra of \ce{Fe(CO)3} and \ce{Fe(CO)4} relative to \ce{Fe(CO)5} simulated using MR-ADC(2)-X with and without spin--orbit coupling (SOC) effects. 
	Vertical lines represent the computed core ionization energies and the corresponding photoelectron intensities.
	Blue lines correspond to \ce{Fe(CO)5}.
	See \cref{sec:ComputationalDetails} for computational details.
	}		
	\label{fig:SOC}
\end{figure}

\cref{fig:SOC} demonstrates the effect of incorporating spin--orbit coupling (SOC) in the MR-ADC simulations of Fe 3p difference XPS spectra for \ce{Fe(CO)3} and \ce{Fe(CO)4}.
The spin--orbit corrections computed at the MR-ADC(1) level of theory are reported in the ESI$\dag$ for each transition metal complex.
Without SOC, the Fe 3p$^{-1}$ states are split by 0.80, 1.02, and 1.23 eV in \ce{Fe(CO)5}, \ce{Fe(CO)4} (C\textsubscript{2v}), and \ce{Fe(CO)3} (C\textsubscript{s}), respectively, due to the interaction with ligand environment. 
Including SOC lifts the degeneracy of the lowest-energy core-ionized state in \ce{Fe(CO)5} and increases the spacing in Fe 3p$^{-1}$ states by 1.21, 1.07, and 1.04 eV for each complex, respectively.
These results suggest that the ligand-field interactions and SOC are equally important in determining the splitting in the Fe 3p$^{-1}$ states and demonstrate that magnitudes of these effects are inversely proportional.
Indeed, neglecting  SOC results in $\sim$ 1.1 eV reduction in the Fe 3p redshift in the simulated XPS spectra (\cref{fig:SOC}).
We note that our MR-ADC predictions of spin--orbit splittings in the iron carbonyl complexes are in a good agreement with the results of MCSCF calculations performed by Leitner et al.\cite{Leitner:2018p44307}
Although both approaches are capable of calculating accurate ionization energies, the MR-ADC method does not require performing separate calculations for the ground and excited states, provides direct access to correlated transition intensities, and does not require including the core or inner-shell valence orbitals in the active space.

\subsubsection{Active Space Selection}
\label{sec:Results:active_space}

\begin{figure}[t!]
	\centering
	\includegraphics[width=1.0\columnwidth]{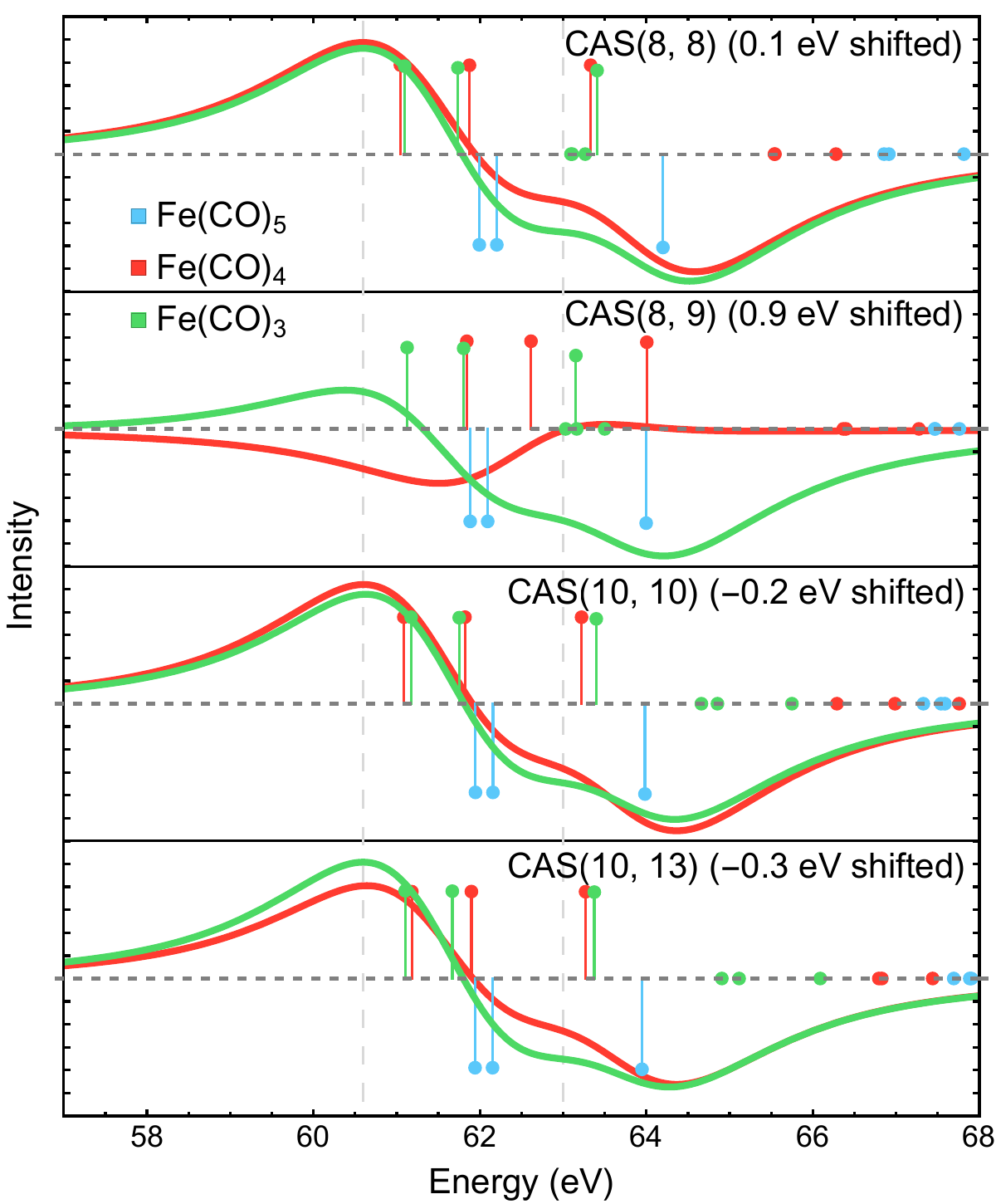}
	\caption{
	Difference \ce{Fe} 3p X-ray photoelectron spectra of \ce{Fe(CO)3} and \ce{Fe(CO)4} relative to \ce{Fe(CO)5} simulated using MR-ADC(2)-X with different selections of active space orbitals in the reference CASSCF calculations.
	Vertical lines represent the computed core ionization energies and the corresponding photoelectron intensities.
	Blue lines correspond to \ce{Fe(CO)5}.
	See \cref{sec:ComputationalDetails} for computational details.
	}
	\label{fig:fe3p_smallAS}
\end{figure}

Our calculations suggest that \ce{Fe(CO)3}, \ce{Fe(CO)4}, and \ce{Fe(CO)5} exhibit predominantly single-reference electronic structures with one electronic configuration accounting for the $\sim$ 80 to 85 \% of the ground-state wavefunction at the CASSCF (10e, 10o) level of theory. 
However, the frontier molecular orbitals of these complexes exhibit significant transfer of electron density from doubly occupied to unoccupied molecular orbitals, indicating the importance of high-order electron correlation effects.
As demonstrated in \cref{fig:FeCO5_CAS} for \ce{Fe(CO)5}, electron correlation is responsible for transferring $\sim$ 0.4 e$^-$ to the higher-lying frontier orbitals, which can significantly change the distribution of valence electrons and screening of core-hole states.
The MR-ADC calculations allow to capture these electron correlation effects, but require selecting an active space.

\cref{fig:fe3p_smallAS} shows the Fe 3p difference XPS spectra simulated using MR-ADC(2)-X with four active spaces ranging from CAS(8e, 8o) to CAS(10e, 13o).
For \ce{Fe(CO)5}, these active spaces are shown in \cref{fig:FeCO5_CAS}, with more details provided in the ESI$\dag$.
The smallest CAS(8e, 8o) active space incorporated four Fe 3d-orbitals and four ligand-based orbitals. 
In CAS(8e, 9o), an additional virtual orbital was added to the active space with the antibonding character between Fe d$_{z^2}$ and axial CO $\sigma$-orbital.
The bonding counterpart of this orbital was added in CAS(10e, 10o).
Finally, three virtual orbitals describing the interaction between $\pi$-antibonding orbitals of CO and Fe 3d orbitals were added in CAS(10e,13o).

The MR-ADC(2)-X calculations with CAS(8e, 8o), CAS(10e, 10o), and CAS(10e, 13o) yield similar spectra that show small differences in relative intensities and peak spacing of features in the transient XPS spectra.
Including the bonding and antibonding combinations of Fe d$_{z^2}$ -- axial CO $\sigma$ orbitals from CAS(8e, 8o) to CAS(10e, 10o) reduces the spacing between the positive and negative features by $\sim$ 0.5 eV, resulting in a closer agreement with the experiment. 
Incorporating the CO $\pi$-antibonding orbitals in CAS(10e, 13o) enhances the intensity of \ce{Fe(CO)3} peaks relative to that of \ce{Fe(CO)4}, which further improves the agreement with the experimental results.
The CAS(8e, 9o) calculations result in qualitatively incorrect difference spectrum for \ce{Fe(CO)4} and significant ($\sim$ 0.9 eV) underestimation in core ionization energies for other complexes. 
These results indicate that both bonding and antibonding Fe d$_{z^2}$ -- axial CO $\sigma$ orbitals are required to properly describe the screening of Fe 3p$^{-1}$ states.
This finding is supported by \cref{fig:FeCO5_CAS} where significant deviations from 2 and 0 are observed in the natural populations of Fe d$_{z^2}$ -- axial CO $\sigma$ orbitals.

\begin{figure}[t!]
	\centering
	\includegraphics[width=1.0\columnwidth]{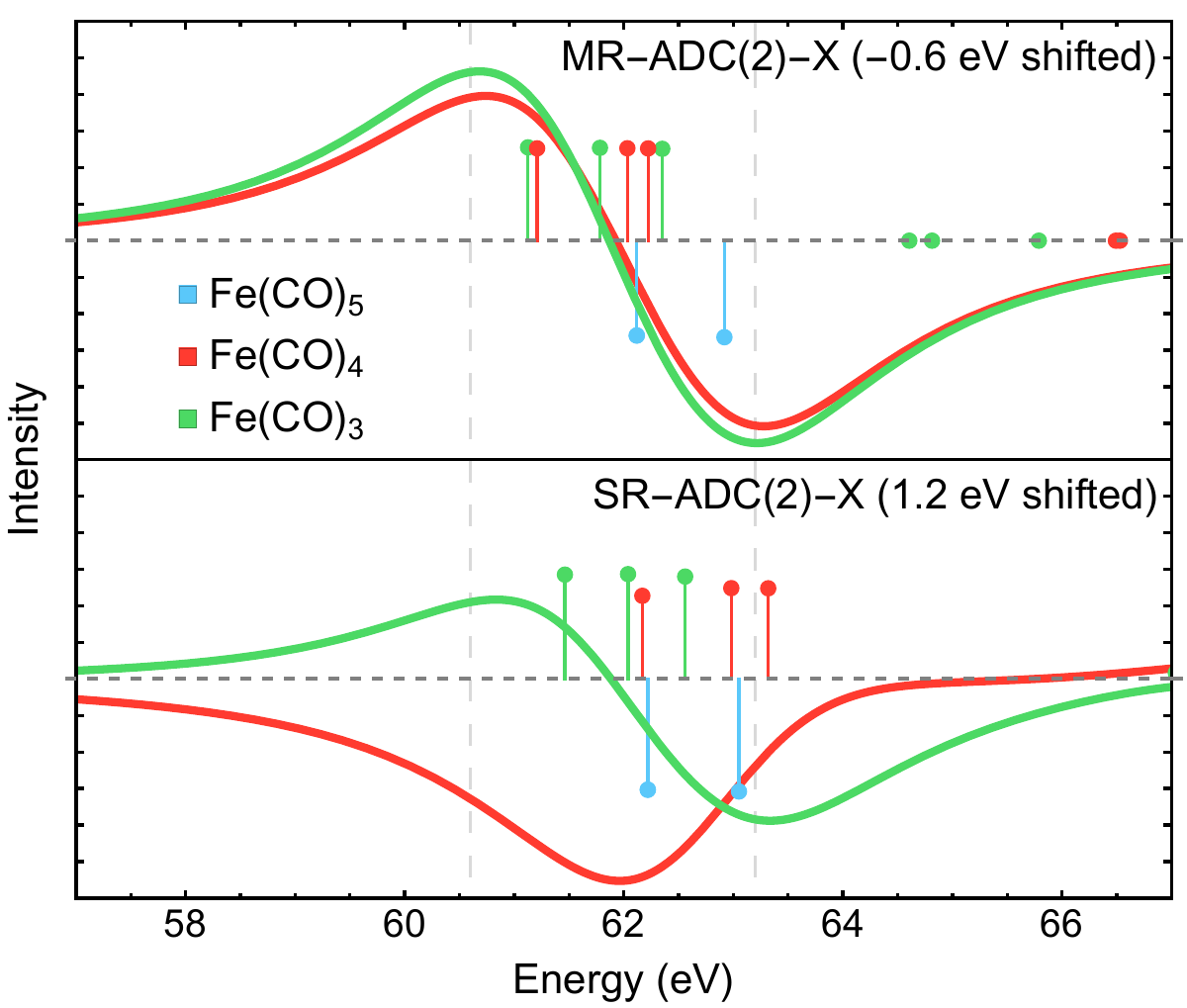}
	\caption{
	Difference \ce{Fe} 3p X-ray photoelectron spectra of \ce{Fe(CO)3} and \ce{Fe(CO)4} relative to \ce{Fe(CO)5} simulated using single-reference (SR-) and multireference (MR-) ADC(2)-X methods.
	Vertical lines represent the computed core ionization energies and the corresponding photoelectron intensities.
	Blue lines correspond to \ce{Fe(CO)5}.
	The simulated spectra did not incorporate the spin--orbit coupling effects and were shifted to align the zero-intensity intercept for \ce{Fe(CO)3}. 
	See \cref{sec:ComputationalDetails} for the information on basis set and active space employed.
	}
	\label{fig:Fe3p_SRvMR}
\end{figure}

To demonstrate the importance of high-order electron correlation effects, we performed the simulations of Fe 3p difference XPS spectra using single-reference ADC(2)-X method (SR-ADC(2)-X)\cite{Schirmer:1998p4734,Dempwolff:2019p064108,Banerjee:2019p224112,Banerjee:2023p3037} without incorporating SOC effects.
As shown in \cref{fig:Fe3p_SRvMR}, SR-ADC(2)-X correctly predicts the spacing between core-ionized states of individual molecules, but does not accurately reproduce the changes in ionization energies between them, predicting almost no redshift in Fe 3p edge between \ce{Fe(CO)4} and \ce{Fe(CO)5}.
In addition, the SR-ADC(2)-X calculations require a larger shift of the difference spectra compared to that of MR-ADC(2)-X in order to reproduce the negative feature relative to the experiment. 

\subsubsection{Basis Set Dependence}
\label{sec:Results:basis_set}

\begin{figure}[t!]
	\centering
	\includegraphics[width=1.0\columnwidth]{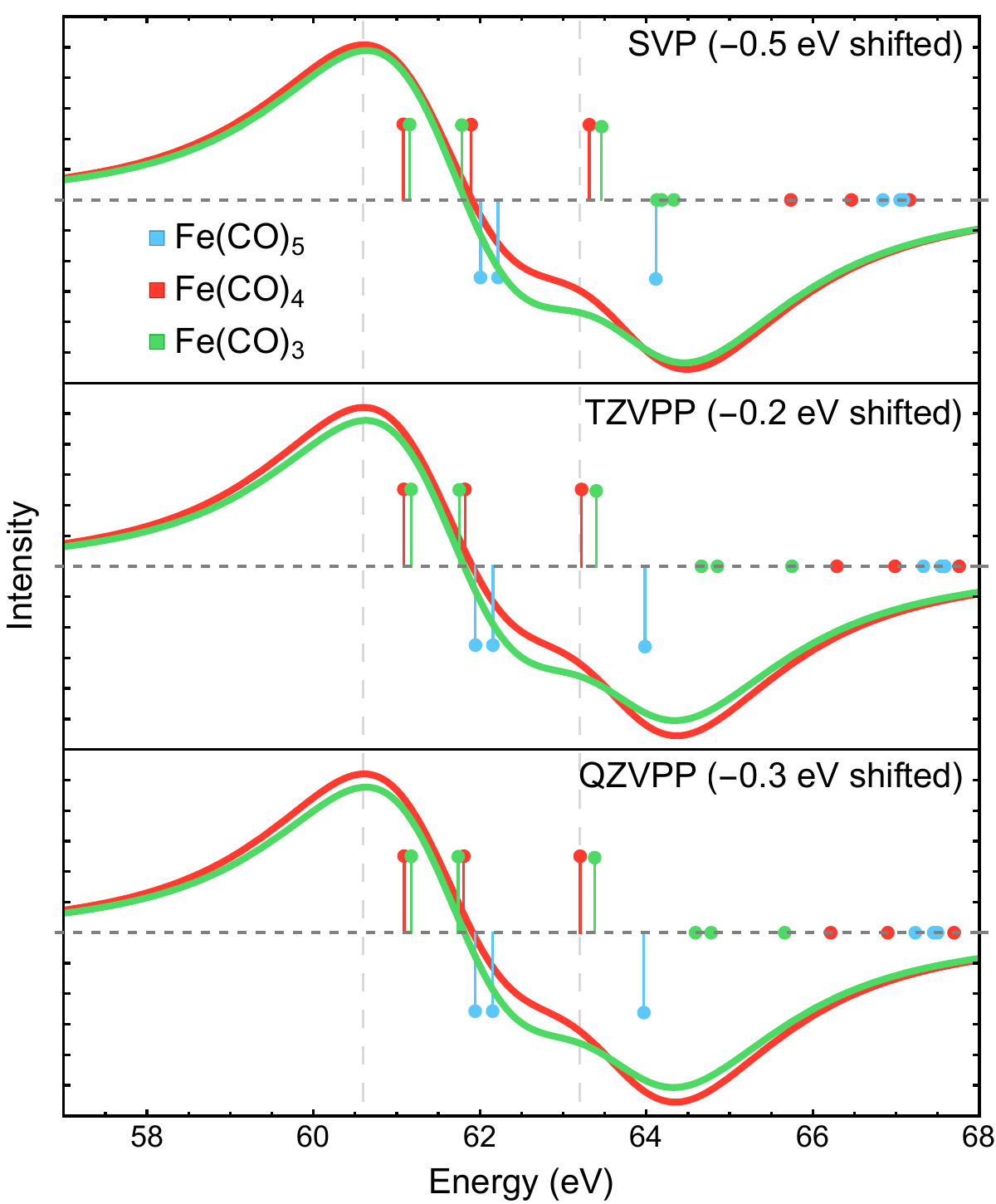}
	\caption{
	Difference \ce{Fe} 3p X-ray photoelectron spectra of \ce{Fe(CO)4} and \ce{Fe(CO)3} relative to \ce{Fe(CO)5} simulated using MR-ADC(2)-X with three fully uncontracted basis sets: def2-SVP, def2-TZVPP, and def2-QZVPP.
	Vertical lines represent the computed core ionization energies and the corresponding photoelectron intensities.
	The simulated spectra were shifted to align the position of zero-intensity intercept with the experimental spectrum.
	Calculations employed the CAS(10e, 10o) active space.
	See \cref{sec:ComputationalDetails} for more computational details.
	}
	\label{fig:fe3p_basissets}
\end{figure}

Finally, we investigate the effect of one-electron basis set on the results of MR-ADC calculations.
\cref{fig:fe3p_basissets} shows the Fe 3p difference XPS spectra simulated using MR-ADC(2)-X with the CAS(10e, 10o) active space and fully uncontracted (unc-) def2-SVP, def2-TZVPP, and def2-QZVPP basis sets.
The unc-def2-SVP spectra overestimate the core ionization energies by $\sim$ 0.2 eV and show small differences in peak spacings relative to the spectra simulated using unc-def2-QZVPP.
The unc-def2-TZVPP results are virtually identical to the unc-def2-QZVPP spectra, apart from $\lesssim$ 0.1 eV shift in ionization energies.
Overall, \cref{fig:fe3p_basissets} demonstrates that the basis set incompleteness errors in the unc-def2-TZVPP calculations performed in this work are expected to be very small.

\section{Conclusions}
\label{sec:Conclusions}

In this work, we performed a computational study of transient X-ray photoelectron spectra (XPS) of \ce{Fe(CO)5}, \ce{Fe(CO)4}, and \ce{Fe(CO)3} for Fe 3p {core-level} and CO $3\sigma$ {inner-valence ionization} using multireference algebraic diagrammatic construction theory (MR-ADC). 
Although our calculations did not incorporate nuclear dynamics effects, the difference XPS spectra simulated using strict and extended second-order MR-ADC methods (MR-ADC(2) and MR-ADC(2)-X) were found to be in a good agreement with experimental results from time-resolved XPS measurements by Leitner et al.\cite{Leitner:2018p44307} where singlet \ce{Fe(CO)4} and \ce{Fe(CO)3} appear as the photodissociation products of \ce{Fe(CO)5} following its excitation with 266 nm light.

Our calculations reveal that the changes in the core-hole screening of Fe 3p$^{-1}$ states due to the loss of CO ligands are not solely responsible for the large ($>$ 2 eV) redshift in the position of Fe 3p feature in the transient XPS spectra.
The observed spectral changes are likely due to a combination of factors that determine the relative energies of the Fe 3p$^{-1}$ states, including strong spin--orbit coupling and significant ligand-field splitting in the core-ionized photodissociated products.
We estimate that the changes in core-hole screening effects amount to only $\sim$ 1 eV redshift in the difference spectra, which is significantly smaller than the splitting of Fe 3p$^{-1}$ states due to spin--orbit coupling and ligand-field interactions.
This analysis suggests that the chemical shifts observed in the transient M-edge XPS spectra of transition metal complexes may originate from several electronic structure effects as opposed to changes in the core-hole screening alone.
In particular, the role of spin--orbit coupling and ligand-field environment must be considered and accurately quantified.
Our results also indicate that, while all three iron carbonyl complexes studied in this work exhibit single-reference electronic structure, high-order theoretical methods must be used to properly account for significant dynamic correlation when simulating their XPS spectra. 

The computational study presented in this work is the first of its kind performed using MR-ADC, which allows to incorporate strong electron correlation in a small subset of frontier molecular orbitals and weaker correlation for the remaining electrons while being able to simulate spectroscopic properties with large one-electron basis sets.
Our results indicate that, in agreement with previous benchmarks,\cite{Moura:2022p4769,Moura:2022p8041} the MR-ADC(2)-X method provides highly accurate results that can be improved systematically by increasing the active space employed in the calculations. 
We also demonstrate that a first-order scheme for treating spin--orbit coupling effects in MR-ADC is highly effective in predicting the splitting of Fe 3p$^{-1}$ energy levels and achieving better agreement with the experimental results.
Further improvements to MR-ADC are underway in our group, including higher-order treatment of spin--orbit coupling and incorporating vibrational effects.

\section*{Conflicts of interest}
There are no conflicts to declare.

\section*{Acknowledgements}
This work was supported by the National Science Foundation (NSF), under Grant No.~CHE-2044648.
Additionally, N.P.G. was supported through the NSF REU grant No.~CHE-2150102.
Computations were performed at the Ohio Supercomputer Center under Project No.\@ PAS1583.
\cite{OhioSupercomputerCenter1987}



\balance




\providecommand*{\mcitethebibliography}{\thebibliography}
\csname @ifundefined\endcsname{endmcitethebibliography}
{\let\endmcitethebibliography\endthebibliography}{}

\end{document}